# Does Quantum Mechanics Violate Bell's Inequality?

**Ghenadie N. Mardari***

*Open Worlds Research*, Baltimore, Maryland, USA

The EPR paradox is known as an interpretive problem, as well as a technical discovery in quantum mechanics. It defined the basic features of two-quantum entanglement, as needed to study the relationships between two non-commuting variables. In contrast, four variables are observed in a typical Bell experiment. This is no longer the same problem. The full complexity of this process can only be captured by the analysis of four-quantum entanglement. Indeed, a new paradox emerges in this context, with straightforward consequences. Quantum mechanics cannot violate CHSH-type inequalities, in the same sense in which local realism cannot do it. The solution is to assume that quantum correlations do not work at the level of full populations, when violations occur, but only apply to incompatible slices of input beams.

## Introduction

The scientific study of physical phenomena is governed by a simple unstated principle: there are no miracles. If something happens, it must be caused by lawful interactions between already existing entities. Yet, Bell's theorem in quantum mechanics appears to defy this principle [1-3]. Real properties can be distributed in many different ways in a population, but they cannot contradict themselves. Therefore, their correlations cannot exceed a predetermined boundary, known as Bell's inequality. Still, quantum property combinations *can* exceed this limit. They seem to emerge either from qualities that do not exist, or from entities that do not interact by direct contact, or from something else no less mysterious [4-6]. Could it be that miracles exist at the quantum level? For all we know, it is possible, but there is a problem. Quantum phenomena cannot be interpreted without contradictions. To give just one example, Bell's Theorem emerged from an attempt to solve the EPR paradox [7]. It lead to numerous advances in quantum technology [8, 9], but it did not remove the conceptual problem it was meant to address. How is it possible for quantum properties to be well-defined and undefined at the same time? Somehow, the formalism of quantum mechanics is self-consistent, the experimental data supports it at every step, but the interpretive concepts do not add up. Perhaps, it is time try something new?

A paradox is usually a symptom of missing knowledge. Either some assumption is incorrect, or some crucial element of analysis is missing altogether, producing contradictions. Though, quantum mechanics is a very complex theory. It has too many "moving parts" to consider. So, instead of looking for solutions from the ground up, the goal of this study was to reverse engineer a quantum paradox. Instead of building a foundation from the known (and potentially insufficient) facts, the strategy was to determine: which elements of the theory *should* be there, for these conceptual problems to go away? The first choice was to begin with the EPR paradox, as the most relevant puzzle for this topic. Surprisingly, this did not pan out as expected. EPR pioneered the study of two-quantum entanglement, and by far the greatest number of Bell experiments were performed with correlated pairs (Refs [10-20] and many others). However, most of these experiments relied on CHSH-type inequalities [21] that study relationships between four variables.

*Contact: gmardari@gmail.com

The full picture of this process cannot be captured with two-quantum observations, because this approach forces the accumulation of various coincidences in isolation from each other. Fortunately, recent advances made it possible to analyze and test quantum behavior in the context of *four-quantum* entanglement [22-24]. These developments enabled a deeper conceptual understanding of Bell-type quantum behavior. As it often happens in quantum mechanics, a new insight lead to a new paradox. This discovery and its implications are presented below. The most important conclusion is that quantum phenomena *can* be described without paradoxes (and without miracles). Oddly enough, this also entails that quantum mechanics *cannot* violate Bell's inequality with population-level statistics. From a logical point of view, existing violations can only be explained by distributions that apply to non-identical wave-function components. This result is verifiable with quantitative analysis, and can also be confirmed by experiment.

**The paradox**

Quantum correlations violate Bell-type inequalities for specially chosen non-commuting variables. This is often quoted as proof that quantum variables cannot have well-defined values prior to measurement [25, 26]. Though, it is important to understand the reason for this: some quantum variables do not have joint distributions [27]. This means they can never belong together to a *whole* population: neither before, nor during, nor after measurement. For an intuitive illustration of this type of property combinations, consider a large population of shirts, measured for three binary properties: *fabric* (cotton/ non-cotton), *color* (white/ non-white), and *sleeve length* (short/ long). Suppose that a measurement procedure yields the following rules of association between the properties:

1. All the cotton shirts are white.
2. All the white shirts are short-sleeved.
3. All the cotton shirts are long-sleeved.

There is an obvious contradiction between these three sets of coincidences. If all the cotton shirts are white and all the white shirts are short-sleeved, then all the cotton shirts should be short-sleeved. The three correlations could not have been registered in the same population. This conclusion would follow even if only a small proportion of shirts (rather than all of them) were found to have contradictory properties. This is what it means to have distributions in violation of a Bell-type inequality. If only a small proportion of objects have incompatible properties, then we have a small violation. If all of them are incompatible, then we have a maximal ("super-quantum" [28-30]) violation. Quantum correlations are proven to obey the so-called Tsirelson bound [31]. Therefore, only a fraction of quanta can have incompatible states.

Another important concept is the difference between physical and statistical compatibility. Non-commuting quantum properties are not physically compatible. They cannot have sharp values in the same context of observation. However, this is neither necessary nor sufficient for statistical incompatibility. For example, a population of objects cannot be split 50-50 and 70-30 at the same time. These operations can only take place in sequence, in different environments (real, or virtual). However, every object in the population is expected to have a well-defined value for both distributions. As long as the same number of objects is subjected to incompatible preparations, the distributions are statistically compatible with each other. In order to achieve a contradiction at the individual level, one must enforce a set of stable conditional relationships between variables. These conditions can only be satisfied without contradiction in separate populations [32]. As seen in the shirt example above, one population must only possess cotton shirts with long sleeves, and the other can only have cottons shirts with short sleeves. Moreover, these conditional relationship must be systemic. Some kind of mechanism must enforce the observed correlations, or else the implied contradictions cannot be verified in a repeatable manner.

In the same vein, Bell violations can only happen if various correlations between pairs of properties are tested independently from each other. This means that every variable of interest must be measured repeatedly. Individual properties can only exhibit contradictions, if they have the opportunity to display alternative states. In a typical Bell experiment (using the CHSH inequality [21]), there are four variables of interest (*e.g.,* **A**, **B**, **C** and **D**), measured in a cyclic arrangement:

$$S = E(\mathbf{A},\mathbf{B}) + E(\mathbf{B},\mathbf{C}) + E(\mathbf{C},\mathbf{D}) - E(\mathbf{D},\mathbf{A}).$$

Every variable is part of two combinations, therefore it must be measured twice. This nuance is often overlooked, because Bell experiments are performed using two-quantum entanglement. In this sort of settings, only two properties can be measured at a time. This limitation *forces* the observers to repeat every measurement, making it seem as if repeated measurements are only a technicality. Shouldn't quantum correlations be the same in any kind of measurement? This is a very important question to ask, because the process of entanglement is not restricted to the two-quantum case. Indeed, four-quantum entanglement has recently become a hot topic of research [22-24], because of its practical advantages in quantum communication protocols. Ergo, it is possible to conduct an experiment in which all the four variables of a Bell test are measured at the same time. Logically, this means that every variable can only have one distribution. At the individual level, one value of variable **A** is used for the coincidence with **B**, and the same value is used for the coincidence with **D**. Yet, how can the same value contradict itself? This is patently impossible. Therefore, such a protocol can never produce violations of any Bell-type inequality.

In light of this conclusion, quantum mechanics appears to contain a new paradox. If quantum variables are measured two at a time with four-quantum entanglement (by ignoring the values of two out of four quanta), then we should expect violations of the CHSH inequality, just like in any other Bell experiment. Yet, such violations are impossible if all the four quanta are measured at the same time, deriving the coefficients of correlation from the record of quadruple coincidences. Both of these predictions must follow with accuracy from the formalism of quantum mechanics, if it is presumed to be a self-consistent theory. Though, how can they take place at the same time? It looks *as if* the rules of association between the values of two quantum variables depend on the observer's decision to measure or not to measure two *other* variables, even if the corresponding quanta are detected at immense distances from each other. This is similar to the EPR paradox for two-quantum entanglement, but this time the ontological implications are directly observable. Instead of an implicit contradiction at the individual level, we have an explicit contradiction at the population level.

### The solution

At first sight, the best way to solve this paradox is by invoking some sort of non-local mechanism. Perhaps, quantum behavior is such that a choice between measuring only two or all four entangled quanta results in different correlations? The problem is that such a hypothesis can only work if it allows exclusively for non-signaling non-locality [33-35], but no such assurance is available in the case of four-quantum entanglement. For instance, it is possible to set up a cosmic experiment, in which two streams of photons are sent to the Moon, while the other two are kept on Earth, in fiber optic loops. Two seconds after launching the quanta, terrestrial observers can decide whether to measure or discard their photons. If non-locality is real, then the open space channels should be instantly affected. Accordingly, Moon-based observers can determine if the remote channels were recorded or not by performing a Bell test with their two streams of photons. This would make it possible to establish superluminal communications between the Earth and the Moon. Ergo, non-local explanations of four-quantum entanglement cannot be compatible with quantum mechanics. There is an instructive precedent for this problem, known as Poppers' experiment [36, 37]. If non-local collapse is assumed to work at the level of single particles, then – it seems – EPR measurements of momentum should display signaling non-locality. Yet, the experiment did not confirm this expectation [36]. More importantly, formal analysis did not support it either [37]. Quantum predictions are derived from the net effect of superposed wave-function components. It is a mistake to interpret them with intuitive concepts about particle behavior. Furthermore, quantum mechanics was built around the correspondence principle [38]. Its predictions for large-$N$ observations (including Bell experiments) cannot contradict well-established macroscopic facts.

The puzzle of four-quantum entanglement can have a physically sound solution, if it has a logical solution. So, we have to inquire: what kind of configuration would make the contradictions go away? It has to be true that pair-wise detection violates Bell-type inequalities. It must also be true that quadruple detection does not violate the same inequalities, in identical experimental settings. Is it possible for both of these observations to take place at the same time, if we are free to make any convenient assumption? Firstly, suppose that the number of double coincidences is equal to the number of quadruple coincidences (meaning that all the quanta are detected in ideal experiments). In this case, it is impossible for both outcomes to be true. Either Bell-type inequalities are violated, or not. The paradox stands. Secondly, suppose that the number of quadruple coincidences is larger than the number of double coincidences. This hypothesis must be dismissed as unsound, because quadruple coincidences also include double coincidences. Finally, consider the possibility that double coincidences outnumber quadruple coincidences. In this case, the paradox vanishes. We can envision an experiment in which a minority of quantum sets produce four coincident events, but most of them produce only two. Bell-type inequalities cannot be violated by the subgroup of quanta that generate quadruple detections. However, every pairwise coincidence (above the four-event threshold) is free from this constraint. If every type of pairwise coincidence belongs to a different slice of the input population, then they can have stable coefficients of correlation. In the special case when paired samples are consistently biased in predictable ways, they can violate the CHSH inequality (or any other Bell-type inequality).

For clarity, quantum experiments are often designed to reveal information about individual modes of propagation from multi-mode input beams. Four-quantum entanglement would be paradox-free if its variables were also defined as properties of wave-function components, rather than whole populations. In particular, one could assume that measurement settings determine which components become observable. If so, then changing these settings may change the subset of detectable quanta. Hence, all the members of an entangled four-quantum set should be detectable if measured in the same way (Fig. 1A). In contrast, if every quantum is measured in a different way, then some members from each group should be likely to miss their detectors, depending on the input component that they represent. For example, some sets might generate coincidences for **A** and **B**, but not for **C** and **D**; others for **B** and **C** only, and so on. To be more specific, some sets could generate only one detection event or none, others would generate two or three coincident events, and only a minority would generate quadruple detections (Fig. 1B). The formalism of quantum mechanics would be entirely self-consistent, if it could predict all of these rates of coincidence with precision. This would explain the ability of quanta to violate Bell-type inequalities, as well as their inability to violate Tsirelson-type inequalities.

An apparent problem for this conclusion is the loophole-free nature of modern Bell experiments [13-18]. How is it possible to have bias-free data and strongly biased samples at the same time? The answer is that quantum predictions cannot apply to whole populations by default, given the arguments above. Therefore, adequate experiments will produce violations without loopholes, because they include component filtering in their basic design. Nonetheless, there is a sense in which quantum predictions require a loophole, because of the way they are presented in the context of debates on local realism. In particular, the fair sampling assumption appears to imply that any input quantum has an equal chance of being detected at the output. Yet, Copenhagen-type interpretations assume that quantum properties are undefined until the moment of collapse, *i.e.*, in the plane of detection. Therefore, the fair sampling assumption is only discussed as an aspect of the detection loophole [13-18, 39]. Presumably, low detector efficiency allows for the possibility of bias in the measurement procedure, weakening the argument against local realism. In other words, the possibility of input-dependent loss at other stages of the experiment (see *Appendix*) is not seen as a problem for the fair-sampling assumption, even though these earlier stages are the ones that would actually matter for a local realist model of quantum behavior. If a detector is 100% efficient, it cannot distort the data, but that is also why it cannot make up for biases that are already built in. Consequently, the argument against local realism is based on circular reasoning. Non-realist and/or non-local assumptions are used to decide which loopholes are relevant, thereby ensuring that the final conclusions are incompatible with local realism.

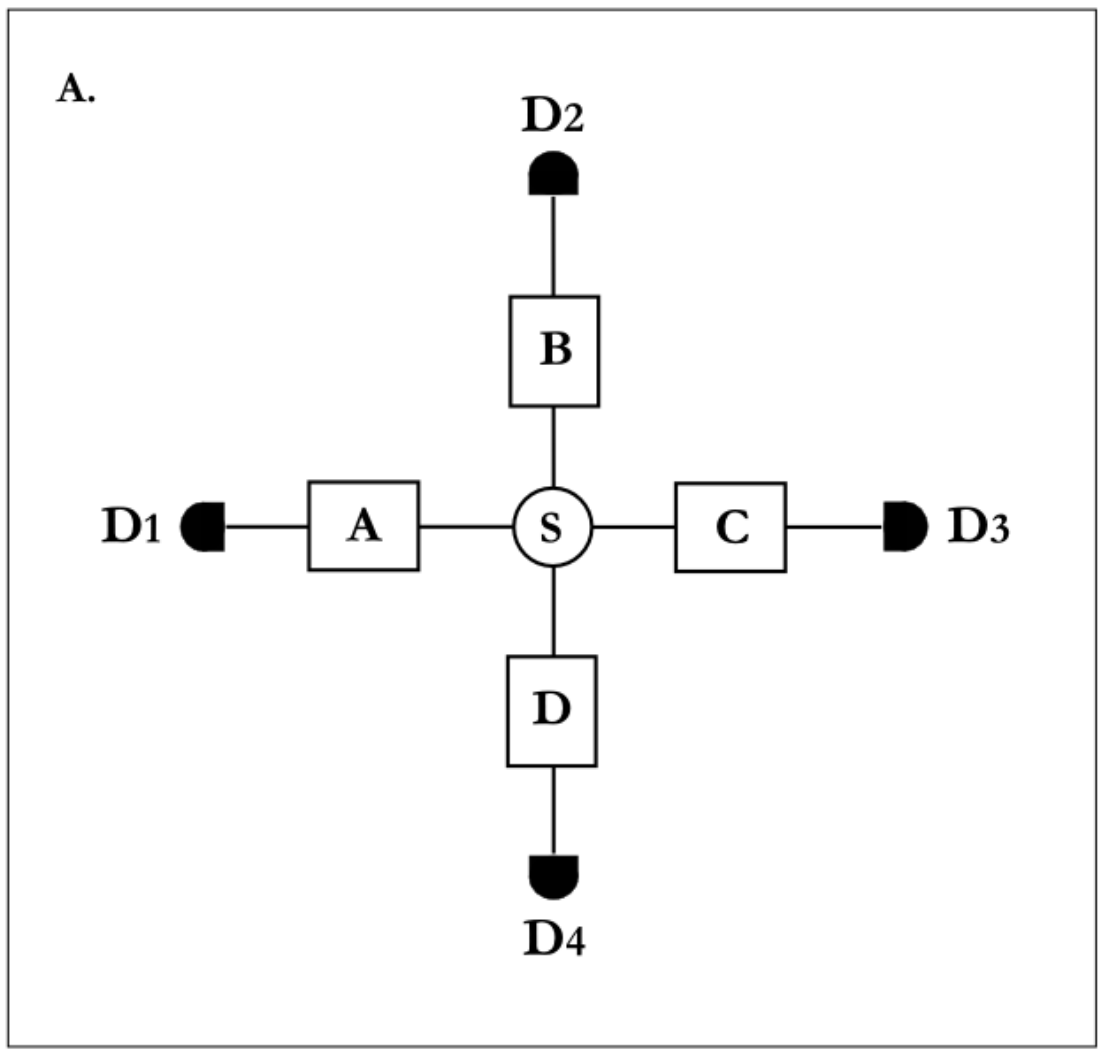

B.

| # | A | B | C | D |
|---|---|---|---|---|
| **1.** | 1 | 1 | 1 | 1 |
| **2.** | x | 0 | 0 | x |
| **3.** | x | x | 1 | 1 |
| **4.** | 1 | x | x | 0 |
| **5.** | 0 | 0 | x | x |
| **6.** | 0 | 0 | 0 | 0 |
| **7.** | 1 | 1 | x | x |
| **8.** | x | 1 | 0 | 1 |

**Fig.1. Four-quantum entanglement without paradox.** **A)** Conceptual diagram. Four entangled quanta propagate in opposite directions from a common (virtual) source-point. The marked boxes (A, B, C, and D) symbolize four different measurement procedures, corresponding to four non-commuting quantum variables. When measured in the same way, all the four quanta have an equal probability of detection. When measured in different ways, these probabilities are individually determined by input properties.

B) Hypothetical data plot, illustrating possible patterns of detection, when Bell-type inequalities are violated without paradox. Values “0” and “1” represent measurement outcomes, while “x” corresponds to missing events. Four-fold coincidences correspond to a minority of trials (in this case, lines 1 and 6), while the majority of trials produce two-fold coincidences, for at least one pair of the variables. This is why different variable combinations can sample incompatible slices of the input population.

## Discussion

One of the central mysteries of quantum mechanics is the phenomenon of superposition. Two detectors, separated by time and space, cannot influence each other in any known way. Yet, they can display sequences of events that emerge from Schrodinger's-cat-style superposition, like overlapping components of a single wave. How can this be? A possible explanation is that space-like separation is an illusion, and that some kind of non-local process is at work. However, the Copenhagen interpretation strongly cautions against any premature conclusions about quantum reality [40]. Detectors extinguish particles, in order to reveal indirect qualities. Therefore, they do not express instantaneous properties of quanta. Furthermore, they cannot expose their past properties either, because quantum distributions are predicted by wave-like processes. Instead, the proportion of quanta at any given coordinate in a distribution reveals information about the amplitude of a corresponding wave-function component. Therefore, quantum observables cannot be used to justify beliefs about the "actual" properties of single particles. Some distributions may look superposed because they reveal information about virtual wave-function components in superposition.

The point of confusion is that wave-like properties can be revealed in two different ways. One way is to expose their input structure, by revealing various components directly. For example, we can have a randomly polarized beam, and isolate input states of polarization, one by one. Another way is to transform the whole wave-function in alternative incompatible ways, by manipulating the relationships between superposed components. In the case of polarization, this entails rotating all the input states to produce only two orthogonal outputs. Which of these two mechanisms is at work in Bell's Theorem? If it is the first one, then Bell-type inequalities can be violated by quantum mechanics, but this cannot serve as a test of realism. If it is the second, then quantum mechanics cannot violate such inequalities. Observing a violation with this method would be proof of signaling non-locality, simultaneously falsifying both classical mechanics and quantum mechanics. That is why it is imperative to review known Bell violations and to determine: do they support existing theories, or do they contradict them? In the same vein, it is necessary to perform Bell experiments with four-quantum entanglement in order to remove any doubt about the mechanism behind quantum correlations in EPR-type experiments.

## Acknowledgments

The final shape of this argument was greatly influenced by discussions with France Čop, Marcus Appleby, Ehtibar Dzhafarov, and Ligia Mardari. I am also grateful to Andrei Khrennikov, Guillaume Adenier, Sören Wengerowsky, Giuliano Scarcelli and Robert Boyd for helpful explanations and references. Finally, credit is also due to the participants of the Växjö Conference series on Quantum Foundations, as well as the SPIE Optics and Photonics series, taking place every summer in Sweden and the USA, respectively.

## Appendix: Understanding Bell-type quantum correlations

The properties of quanta are very hard to interpret, because there are too many models with competing assumptions and rules of analysis. The best strategy for progress in this environment is to sidestep all the subtleties and to start directly with a "bare-bones" example. As shown in Fig.2, optical states of momentum and position are determined by passing a beam through a lens [41]. Here, a multimode beam of quanta is emitted from the source plane (*Sp*), with only two source-points shown, for clarity. It propagates for a while and then it is distorted by the lens. This distortion causes the beam to evolve through a new sequence of profiles, including a sharp momentum spectrum in the focal plane of the lens (*Fp*), and a sharp position spectrum in the image plane (*Ip*). Let us focus on the measurement of momentum first. All the rays with parallel directions converge onto a single spot. Therefore, all the photons that land on the same coordinate in the focal plane correspond to a single state of momentum. This is why the spectrum of momentum is sharp in this plane of observation. Yet, the big question is: what can we learn about the actual properties of detected particles in this experiment? We know that they correspond to a well-defined momentum component, but do we know the instantaneous momentum of individual quanta? As shown in the referenced diagram, the answer has to be negative. The input rays are parallel before they reach the lens, but they are all refracted so as to converge on focal points. Therefore, the momentum profile at the moment of detection is undefined. Only the information about the input state is well-defined. Though, does this mean that we know the input state of individual particles? No, again. The proportion of quanta at a given point of detection, relative to the whole distribution, confirms the amplitude of the corresponding component in the input wave-function. Still, individual quantum histories (real or imaginary) remain shrouded in uncertainty. We know that quanta have a fixed probability of landing at the chosen point, but we cannot blindly assume they follow rectilinear trajectories from the source to the detector. Optical rays are only meant to capture the geometrical structure of electromagnetic beams. Thus, quantum observations cannot be used to say anything certain about the real properties of detected quanta. Instead, they seem to reveal information about governing wave-functions in general, and their superposed components in particular. This conclusion applies in equal measure to measurements in the image plane, where the so-called position information can be resolved.

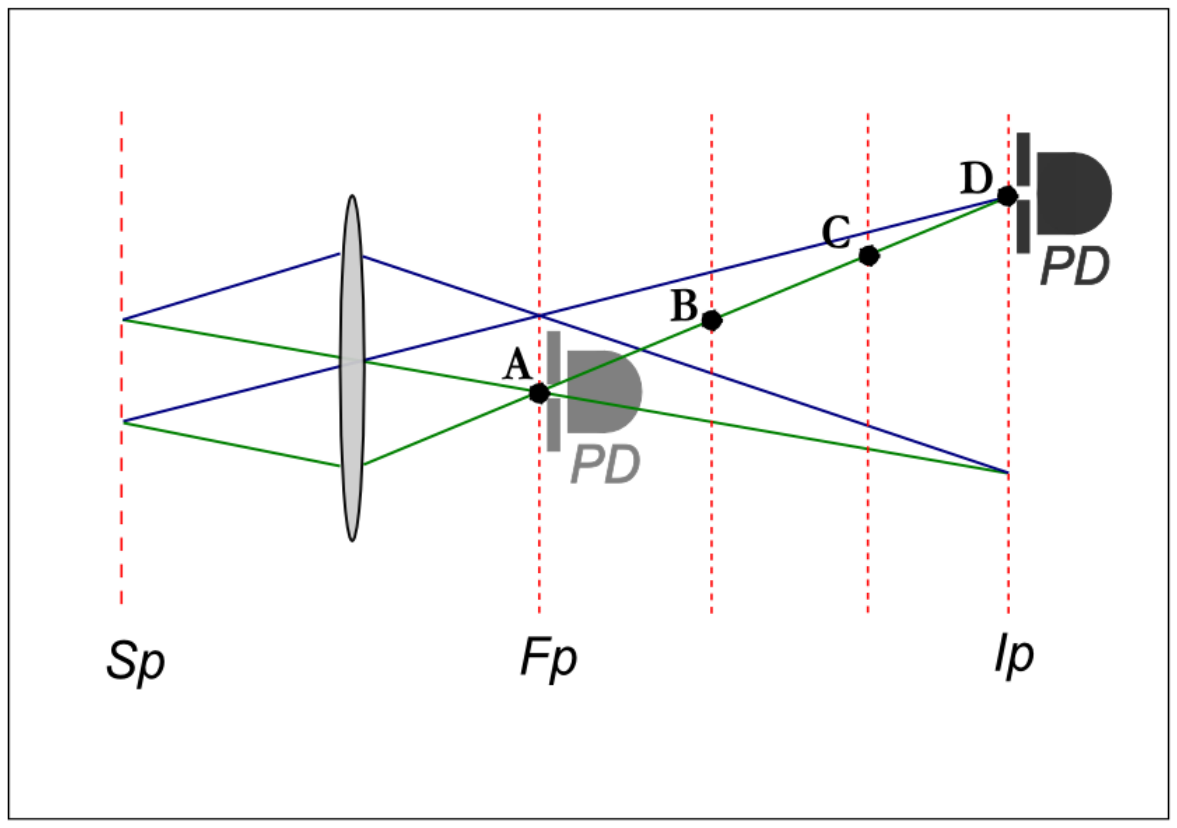

**Fig. 2. Understanding Bell-type variables.** A multimode beam is propagating from the source plane (Sp) towards the image plane (Ip). Only two source points with two optical rays each are shown, to highlight the geometrical structure of the corresponding wave-function. Quantum measurements isolate single points from wide planes of detection. In this case, photonic momentum is determined by placing a pinhole detector (PD) in the focal plane of a lens. Position information is obtained by moving the same detector to the image plane. A Bell-type experiment can be performed in this arrangement by adding more planes for observation. Even if all the points of detection (A, B, C, and D) are connected by a single optical ray, the samples of coincident photons for each pair of variables must remain incompatible. Due to the Uncertainty principle, every quantum has a predictably limited probability of passing through two points in a row. Many of the photons that cross points B and C will be different from the photons that cross points A and D.

As a corollary of the above, quantum properties are fundamentally non-classical. Strictly speaking, they do not belong to the objects under investigation. Nonetheless, every quantum gets a value tag at the detection, and this is the only thing that matters for statistical analysis. As seen in Fig. 2, quanta can be detected in various planes. Every one of these planes can be described as a continuous variable. In principle, every quantum can be attributed a sharp value for each plane, producing a quasi-classical data-plot. For example, it is possible to create EPR states, using one quantum for a measurement of momentum, and another for a measurement of position [41]. As long as all the quanta are measured, and all the values are attributed to unique recipients, the outcome is a joint distribution. Ergo, such properties cannot violate Bell-type inequalities. Then again, how can we get predictions *with* violations, and without paradoxes? From an interpretive point of view, quantum properties are better captured by continuous variables. However, actual experiments require point-like detectors, in order to isolate individual states from each other. Therefore, continuous variables have to be converted into dichotomous variables. Every value in a continuous variable is redefined as a binary quality (*e.g.,* with a value of “1” if a quantum produces a click at the chosen position, and a value of “0” if it does not). For example, a distribution of momentum with 12 values can be redefined by choosing a single point of detection, such that one value (out of the original 12) becomes “1”, and the remaining 11 become “0”, because they cannot produce clicks. If it was possible to still count all the quanta in this context, we would get joint distributions again. Unfortunately, undetected quanta cannot be counted. The number of detected events is dramatically reduced. For instance, if 12 equiprobable states of momentum and 12 equiprobable states of position are recodified into a pair of binary variables, then the number of joint events for two dichotomous variables is 1/144 from the total number of quanta. Thus, we are no longer dealing with a relationship between two population-level variables. Instead, we are dealing with a relationship between two values (corresponding to two incompatible wave-function components) extracted from a pair of originally continuous variables.

This experimental setting can be expanded into a Bell-type *gedanken* experiment, by choosing to detect quanta in two intermediate planes, in addition to the original two (focal and image). This would raise the number of binary variables to four, with instructive consequences. If four entangled quanta are used, each of them can be detected in a different plane. However, as mentioned above, quantum trajectories are uncertain (and essentially chaotic). Therefore, it should be impossible to find a line that connects four points in the chosen planes and also results in detection events for all and the same photon ensembles (see Fig.2). Of course, it is possible that some entangled sets will click at all the four points of observation. However, it is at least as likely that other sets will not. Indeed, the probability for each occurrence could be derived by taking into account the profile of the wave-function, as it evolves from one plane of detection to the next. The important implication is that most quadruplets will not produce four-event coincidences. This means that incompatible coefficients of correlation are likely to be displayed, in violation of the CHSH inequality. Still, these violations must be limited by the subset of quanta that *do* produce four-fold events, in agreement with Tsirelson’s theorem. The best way to describe this outcome is that: “*Every A+ is a B+*”, “*Every B+ is a C+*”, but “*Some A+ are not C+*”. From a logical point of view, this set of rules is consistent, because “*every X+*” is defined as shorthand for “*every X+ that coincides with Y+*”. (So, it does not really include *every* “X+”). For instance, different subsets of “A+” single events may coincide with different alternative events (“B+” or “D+”). In short, the extent of Bell violations can be predicted entirely by the incompatibility between the types of samples that can produce coincidences in various combinations. Accordingly, the success of a Bell experiment depends on the ability of the observer to isolate the necessary wave-function components and to filter out the rest of the input beam.

Quantum measurements of polarization are essentially similar to quantum measurements of position and momentum. In particular, randomly polarized beams can be assumed to contain many superposed components, just like chaotic beams with randomly distributed wave vectors. Still, this similarity is obscured by two factors. Firstly, quantum states of polarization can only be revealed in orthogonal pairs. This is a disadvantage, because it is not possible to capture the full spectrum of the input profile with one device, as it can be done with a lens for momentum/position states. Hence, it is not clear: do all the possible component states exist at once, or only the orthogonal pairs, in sequence? Secondly, measurements of polarization are automatically dichotomous, because they result in two output projections, transmitted and reflected. This makes it possible to count both types of values ("0" and "1"), in contrast to momentum measurements that only register one type. Does this entail that all the quanta are detectable in Bell experiments with photons? It may seem tempting to believe so, but this is logically impossible. As shown above, it does not matter if the underlying physics is classical or non-classical. If all the members of a population are detected, then all the value tags are distributed in compatible ways. This means that measured variables have joint distributions and cannot violate Bell-type inequalities. Thus, quantum measurements of polarization must be similar to quantum measurements of momentum, producing information about isolated wave-function components. (Quantum mechanics would be inconsistent, if they were not). The difference can only be quantitative, in that two individual states (out of many) are observed, instead of just one. In other words, quanta with nearly vertical states of polarization must have a high probability of being transmitted; quanta with nearly horizontal states of polarization must have a high probability of being reflected; while quanta with nearly diagonal states of polarization must have a high probability of being discarded. Rotating a polarizer would give higher probabilities of coincident detection to different input components, resulting in incompatible samples and corresponding Bell violations. For example, some photons can be presumed to have objective states of polarization between 0° and +22.5° at the input. Others could be presumed to be aligned between 0° and -22.5°. Therefore, the first group has a higher probability of generating coincident events for measurements at 0° and +22.5°, while the second group would be at a disadvantage. Simultaneously, a similar mechanism could be at work in the reflected channel, for measurements at 90° and 112.5°.

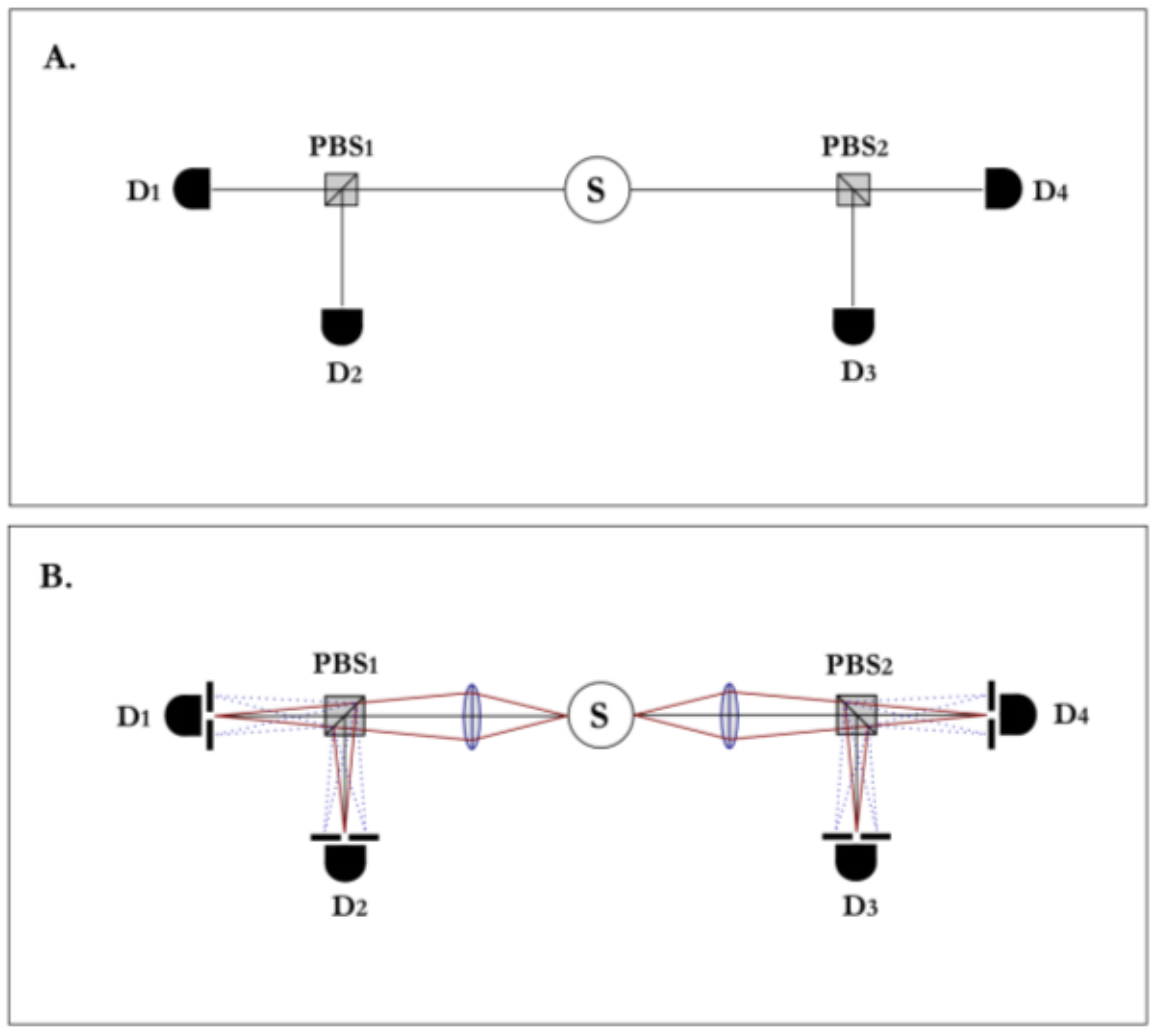

**Fig. 3. Two kinds of Bell violations.** As argued in the text, the goal of quantum experiments is to produce information about the components of input beams. In the case of polarization measurements, this can be achieved in two ways. **A)** Experiments with dim polarizers. Birefringent beam-splitting crystals allow the horizontal and vertical states of polarization to pass unperturbed. In contrast, they distort intermediate states of polarization. If diagonal components are absorbed at a higher rate, then Bell-type inequalities can be violated. Every time the beam-splitters are rotated, different samples of photons are allowed to produce coincident events.
**B)** Experiments with bright polarizers. Birefringent crystals with low absorbency cannot have the same effect. However, they induce phase changes for the diagonal components that pass through. As a result, affected quanta are more likely to drift away from the target of detection in focused projections. That is why Bell-type inequalities can still be violated, if a lens is added to the experimental set-up, in combination with a pinhole detector (or single mode optical fiber).

Are there reasons to expect such a process of discrimination in actual quantum experiments? Indeed, there are two possible mechanisms (Fig. 3). Firstly, some polarizers are physically unable to pass incoming radiation in full. In many cases, they absorb more than 50% of input projections (as seen, for example in early Bell experiments in free space [10], or more recent experiments with wave-guide beam-splitters [20]). Yet, birefringent crystals do not have homogenous effects. If an input state of polarization is parallel to one of the two optical axes (ordinary and extraordinary), it can pass unaffected. In contrast, diagonal inputs cannot pass through without disturbance. Is it not reasonable to suspect that perturbed components are absorbed more often? Secondly, some polarizing cubes are highly transparent, with negligible rates of absorption. Yet, quantum experiments with such beam-splitters require additional elements, such as a lens in combination with a pinhole detector (or a single-mode optical fiber) [17-19]. This opens up a different avenue for polarization-dependent quantum loss (Fig. 3B). When birefringent crystals encounter diagonal states of polarization, they must rotate them towards the fast (ordinary) axis, or towards the slow (extraordinary) axis. This results in phase changes for the corresponding components. As it is known from classical optics, focused projections emerge from the superposition of Huygens-Fresnel components, when lenses induce appropriate phase-delay distributions [42]. (In the field of quantum imaging, this is also described as Fresnel propagation for probability waves [37]). If the relative phase between some components is changed, the focal point drifts. Therefore, diagonal input components have a higher probability of drift, and a smaller probability of landing on the point of detection, compared to their horizontal or vertical counterparts. To be clear, the specified optical elements in Bell experiments are used for other explicit purposes (*e.g.,* to isolate single modes, to remove noise, or to increase the rates of coincidence). Still, it is remarkable that successful Bell experiments require these elements, when beam-splitter absorption rates are low, just like they logically require polarization-dependent quantum loss to avoid paradoxes. These technical details may seem irrelevant for quantum models with non-local collapse, but it is precisely the goal of basic experiments to test the assumptions behind such models, especially when they lead to paradoxes.